\documentclass[preprint,prd,nofootinbib,tightenlines,amsmath]{revtex4}
\usepackage[latin1]{inputenc}
\usepackage{bbm}
\usepackage{bm}
\usepackage{graphicx}
\usepackage{epsfig}
\usepackage{subfigure}
\usepackage{latexsym}
\usepackage{amsmath}
\usepackage{amsfonts}
\usepackage{amssymb}
\usepackage{color}
\usepackage{hyperref}

\newcommand{\be}{\begin{eqnarray}} 
\newcommand{\ee}{\end{eqnarray}}

\newcommand{\bmp}{\noindent\begin{minipage}{16cm}}
\newcommand{\emp}{\end{minipage}\vskip 7mm} 
\def\lsim{\mathrel{\raise.3ex\hbox{$<$\kern-.75em\lower1ex\hbox{$\sim$}}}}
\def\gsim{\mathrel{\raise.3ex\hbox{$>$\kern-.75em\lower1ex\hbox{$\sim$}}}}

\begin{document}

\baselineskip=15pt

\hspace*{\fill} $\hphantom{-}$

\title{Resizing the Conformal Window: A $\beta$-function Ansatz}
\author{O.~Antipin}
\email{oleg.a.antipin@jyu.fi}
\affiliation{Department of Physics, University of Jyv\"askyl\"a, P.O.Box 35, FIN-40014 Jyv\"askyl\"a, Finland 
\\
and 
Helsinki Institute of Physics, P.O.Box 64, FIN-00014 University of Helsinki, Finland}
\author{K.~Tuominen\footnote{On leave of absence from Department of Physics, University of Jyv\"askyl\"a}}
\email{kimmo.tuominen@jyu.fi}
\affiliation{
CP$^3$-Origins, Campusvej 55, 5230 Odense, Denmark.\\
and Helsinki Institute of Physics, P.O.Box 64, FIN-00014 University of Helsinki, Finland\\}

\date{\today}

\begin{abstract} 
\vspace{2mm}
We propose an ansatz for the nonperturbative beta function of a generic non-supersymmetric Yang-Mills theory with or without fermions in an arbitrary representation of the gauge group. While our construction is similar to the recently proposed Ryttov-Sannino all order beta function, the essential difference is that it allows for the existence of an unstable ultraviolet fixed point in addition to the predicted Bank-Zaks -like infrared stable fixed point. Our beta function preserves all of the tested features with respect to the non-supersymmetric Yang-Mills theories. We predict the conformal window identifying the lower end of it as a merger of the infrared and ultraviolet fixed points. 
\end{abstract}

\pacs{PACS numbers: }

\maketitle

\section {Introduction}

Infrared conformal field theories have received a lot of attention recently in high energy physics thanks to the progress made in understanding the phase diagram of strongly coupled gauge theory as a function of number of colors and flavors as well as matter representations \cite{Sannino:2004qp}. Phenomenologically, (quasi)conformal field theories have application in development of models for dynamical electroweak symmetry breaking \cite{Dietrich:2005jn}, for unparticles \cite{Georgi:2007ek} and the associated LHC phenomenology. In these cases one is interested in theories which are (quasi)conformal in the infrared, i.e. the $\beta$-function of the theory has a non-trivial zero at finite value of the coupling, $\beta(g^\ast)=0$. 

Perturbatively very little is known of the existence and properties of such conformal theories. The Banks-Zaks analysis \cite{Banks:1981nn} for the fixed point is valid typically only near the boundary where asymptotic freedom is lost and is unjustified for physically viable numbers of colors and flavors. Similar uncertainties arise in the analysis aimed to determine the lower boundary of the conformal window using truncated Schwinger-Dyson equations (ladder approximation) \cite{Appelquist:1988yc,Cohen:1988sq}. However, these methods serve as important tools to gain at least qualitative intuition of the full dynamics of strongly coupled gauge theories. Using such methods, it was concluded in \cite{Sannino:2004qp} that phenomenologically interesting theories could be constructed with relatively small new matter content if higher representations with respect to the new strong dynamics were utilized. Several candidate theories for walking technicolor model building were identified. Recently these models have been investigated also on the lattice. For two flavors in the adjoint representation of the SU(2) gauge group, using unimproved Wilson fermions, it has been argued that the theory has an infrared fixed point \cite{Hietanen:2008mr,Catterall:2007yx}. For two flavors in the sextet representation of the SU(3) gauge group it was first argued that there also a fixed point exists \cite{Shamir:2008pb}, but as properly improved Wilson fermions were implemented into the simulations the conclusion is weaker as the theory appears to be more on the sill of the conformal window \cite{DeGrand:2008kx}. Lattice calculations have also been performed for SU(3) gauge theory with fundamental fermions, and these calculations support the conclusion that for $N_f=12$ the theory has a conformal fixed point in the infrared \cite{Appelquist:2007hu}.

Given all this, it would be highly desirable to know the $\beta$-function for a generic SU(N) gauge theory with matter. This has been achieved in supersymmetric cases where supersymmetry provides enough constraint to allow for analytic progress. For ${\mathcal{N}}=4$ super Yang--Mills (SYM) the $\beta$-function is identically zero while for ${\mathcal{N}}=1$ super-QCD (SQCD) the exact $\beta$-function has been determined \cite{Novikov:1983uc}. In particular it has been shown that for $3N_c/2<N_f<3N_c$ SQCD has an infrared stable fixed point \cite{Seiberg:1994pq,Intriligator:1995au}. An important progress for non-supersymmetric theories was made in \cite{Ryttov:2007cx} where an ansatz generalizing the SQCD $\beta$-function for non-supersymmetric gauge theories was introduced. This ansatz was constructed to reproduce known perturbative results and to match on its supersymmetric counterpart in appropriate limits. Furthermore, this $\beta$-function ansatz was applied to determine the conformal window, i.e. the numbers of colors and flavors for which the non-supersymmetric SU($N_c$) gauge theory with fermions in a given representation has an infrared stable fixed point. Regarding this question, consider  \cite{Kaplan:2009kr} where it has been argued that there are three generic mechanisms how conformality is lost in quantum theories. Namely, 1) the fixed point can go to zero coupling, 2) it can go to infinite coupling or 3) the Infrared fixed point (IRFP) may annihilate with an ultraviolet fixed point (UVFP) whence they both disappear into the complex plane.

Of these mechanisms 1) and 2) are afforded by SQCD. To see this, recall the result of  Seiberg~\cite{Seiberg:1994pq,Intriligator:1995au} that SQCD is conformal in the window $3/2 \le x\equiv N_f/N_c \le 3$ where $N_c$ $(N_f)$ is number of color (flavors).  For $x$ just below three, the theory has a Banks-Zaks fixed point at weak gauge coupling $g$ \cite{Banks:1981nn}, and approaching $x=3$ from below, this fixed point merges into the trivial fixed point at $g=0$. For $x>3$ the theory is in the asymptotically non-free phase, a non-Abelian free electric phase, and the phase transition from $x\lsim 3$ for which the theory has a fixed point to $x>3$ whence the fixed point does not exist anymore provides an example of mechanism 1) above.  In contrast, then, approaching the lower end of the conformal window at $x=3/2$ from above, SQCD undergoes a phase transition from a strongly coupled conformal theory when $x\gtrsim 3/2$  to yet another different phase. The nature of this phase is most straightforwardly exhibited by resorting to the electric-magnetic duality: The above discussion of the phases has been explicitly carried out for the electric degrees of freedom. However, for SQCD there is an equivalent description in terms of the magnetic degrees of freedom which are composites of the elementary electric degrees of freedom. The magnetic sector is described by SU($N_f-N_c$) gauge theory which is not asymptotically free for $N_f\lesssim 3/2N_c$ and hence is weakly coupled at large distances; the theory is in the non-Abelian free magnetic phase. Since the electric and magnetic couplings are related as $g_M\sim 1/g_E$, what happened in terms of the magnetic degrees of freedom above corresponds to the loss of  conformality in the electric description via mechanism 2). Hence the scenarios 1) and 2) can describe the same physics in terms of different (magnetic or electric) degrees of freedom. Of course this needs not to be the case unless one can establish an electric-magnetic duality as in the example above. An interesting development for the QCD dual, i.e. an attempt to generalize the "Seiberg program" used in the above discussion to non-supersymmetric cases, appeared recently \cite{Sannino:2009qc}.
  
However, it is interesting to note that mechanism 3) seems not to be allowed either with the $\beta$-function of SQCD or with the $\beta$-function ansatz of \cite{Ryttov:2007cx} for non-supersymmetric SU($N_c$) gauge theory. On the other hand, it was shown in  \cite{Kaplan:2009kr} that mechanism 3) is realized in a wide class of non-supersymmetric theories. Quantitative indications for this phenomena in QCD with many flavors were presented in \cite{Gies:2005as}. Along these lines, it was also discussed in \cite{Kaplan:2009kr}, that in SU(3) gauge theory with fermions in the fundamental representation QCD-like vacuum with spontaneously broken chiral symmetry arises at large $N_c$ and $N_f$ when Banks-Zaks fixed point annihilates with another UVFP as $x=N_f/N_c=x_{\rm{crit}}$. It was therefore predicted that above this critical value $x_{\rm{crit}}$, SU(3) gauge theory would possess an UVFP called QCD$^*$  (in addition to IRFP and the free UV fixed point). In an attempt to identify QCD$^*$ theory several models were considered but no theory possessing the full chiral symmetry of QCD in the perturbative regime was found leaving the value of $x_{\rm{crit}}$ unknown.

In this paper we will combine the ideas of \cite{Ryttov:2007cx} and \cite{Kaplan:2009kr} and propose an alternative ansatz for an all order $\beta$-function for a generic Yang--Mills (YM) theory featuring the mechanism 3). Having an analytic proposal will enable us to draw concrete numerical predictions uncovering rich dynamics. Closing the parallel with \cite{Kaplan:2009kr}, we will consider QCD$^*=$QCD which means that YM gauge coupling has an {\it additional} unstable UVFP at the strong coupling and this will allow us to predict the value of $x_{\rm{crit}}$.

To introduce notations we will first review the Ryttov-Sannino $\beta$-function proposal  \cite{Ryttov:2007cx} in Sec.\ref{all} and then move on to present the modified ansatz in subsequent sections discussing first the large $N_c$ limit and then phenomenologically including the ${\mathcal{O}}(1/N_c)$ corrections. We will compare the conformal windows which we obtain with the existing literature in Sec. \ref{conclusion} where we also present our conclusions and outlook.

\section {Ryttov-Sannino $\beta$-function proposal}\label{all}

As we review the basic results of \cite{Ryttov:2007cx} we also redefine some notations to facilitate smooth comparison with \cite{Kaplan:2009kr}. Define the rescaled 't Hooft coupling
\be
	a\equiv \frac{g^2N_c}{(4\pi)^2},
\ee
and denote $N_f/N_c\equiv x$. In the perturbative $\beta$-function for a generic non-supersymmetric YM theory with $N_f $ Dirac fermions in a given representation $R$ of the gauge group, two lowest order coefficients are universal, i.e. independent on the renormalization scheme used to compute them:
\be
	\beta (a) &\equiv & \mu\frac{da}{d\mu}\nonumber\\
	&=& -\frac{2}{3}\overline{\beta}_0(x)a^2 - \frac{2}{3}\overline{\beta}_1(x)a^3 \ ,
	\label{perturbative}
\ee
where $\mu$ is the renormalization scale and the first two $\beta$-function coefficients are given by
\be
	\overline{\beta}_0(x) &=&{11}\frac{C_2(G)}{N_c}- {4}T(R)x \\
	\overline{\beta}_1(x) &=&{34}\frac{C_2^2(G)}{N_c^2} - {20}\frac{C_2(G)}{N_c}T(R) x  - 12\frac{C_2(R)}{N_c} T(R) x  \ .
\ee
We are using standard group theory notation, with generators $T(R)^a,\, a=1\ldots N^2-1$ of the gauge group in the representation $R$ normalized according to $\text{Tr}\left[T(R)^aT(R)^b \right] = T(R) \delta^{ab}$. The quadratic Casimir $C_2(R)$ is given by $T(R)^aT(R)^a = C_2(R)\mathbbm{1}$ and the trace normalization factor $T(R)$ is related with quadratic Casimir via $C_2(R) d(r) = T(R) d(G)$ where $d(R)$ is the dimension of the representation $R$. The adjoint representation is denoted by $G$. These group theory factors for the representations used in this paper can be found explicitly in \cite{Ryttov:2007cx}. It is useful to introduce a further variable
\be
	X\equiv 2T(R)x=2T(R)\frac{N_f}{N_c}
\ee
which for the fundamental representation is equal to $x$.

The generic form of the $\beta$-function ansatz in \cite{Ryttov:2007cx} was motivated by two features. First, that the perturbative expression for the anomalous dimension $\gamma$ of the quark mass operator,
\be
\label{gamma_pert}
	\gamma(a) = 6\frac{C_2(R)}{N_c} a + {\mathcal{O}}(a^2) \ ,
\ee
depends on $C_2(R)$ which also appears explicitly in the last term of the second coefficient of the $\beta$-function. Second, with the notation defined above, the exact $\beta$-function of SQCD \cite{Novikov:1983uc} takes the form
\be
	\beta (a) &=& - 2a^2
	\frac{\overline{\beta}_0(X)+X\gamma(a)}{1-2a} \ , \\
	\gamma(a) &=& - 4a\frac{C_2(R)}{N_c} + {\mathcal{O}}(a^2) \ ,
\ee
where $\gamma (a) = -\frac{N_c}{(4\pi)^2}d \ln Z(\mu) /d \ln \mu $ is the anomalous
dimension of the matter superfield and $\overline{\beta}_0(x) = 3 - X$ 
is the first coefficient of the perturbative $\beta$-function.

The proposal of \cite{Ryttov:2007cx} was then to write all-order $\beta$-function in the following form:
\be
\label{unmod}
	\beta(a) &=&- \frac{2}{3}a^2 \frac{\overline{\beta}_0(X) - X \,
	\gamma(a)}{1- \frac{2aC_2(G)}{N_c}\left( 1+ \frac{6\overline{\beta}_0^\prime(X)}{\overline{\beta}_0(X)} \right)} \ ,
\ee
with $\overline{\beta}_0(X)=11\frac{C_2(G)}{N_c}-2X$ and
\be
	\overline{\beta}_0^\prime (X) &=& C_2(G)/N_c - X/2  \ .
\ee
For the group SU($N_c$) we have $C_2(G)=N_c$ which simplifies the above equations in this case; however, the form which we have written can be applied for other gauge groups as well. It is easy to show that the above $\beta$-function reduces to Eq. (\ref{perturbative}) when expanding to ${\mathcal{O}}(a^3)$. Since only the two-loop $\beta$-function has universal coefficients, it was assumed in \cite{Ryttov:2007cx} that there is a scheme in which the proposed $\beta$-function is complete.

Decreasing the number of flavors slightly from the point of the loss of asymptotic freedom where $\overline{\beta}_0=0$, one expects a perturbative zero in the $\beta$-function to occur \cite{Banks:1981nn}. Using Eq. (\ref{unmod}) the analysis of \cite{Ryttov:2007cx} near this IRFP shows that at the IRFP the denominator is positive and remains finite as $\overline{\beta}_0=0$ point is approached. Due to existence of the perturbative IRFP, analysis was extended to a lower number of flavors and the critical number of flavors below which the unitarity bound \cite{Mack:1975je,Flato:1983te,Dobrev:1985qv} $\gamma=2$ is violated was used as a bound on the lower end of the conformal window. It was further shown that $\beta$-function in Eq. (\ref{unmod}) reproduces the exact $\beta$-function for $\cal{N}$=1 SYM.  In addition to that, pure YM case was analyzed in detail. The deviation of the $\beta$-function ansatz from the perturbative two-loop result was presented and compared to the deviation of the lattice data with respect to the same two-loop result. The size of the corrections in both cases were found to be of similar magnitude.

As discussed in the introduction, to gain more insight on how conformality is lost in non-supersymmetric gauge theories as a function of number of colors and flavors, we will next modify Eq. (\ref{unmod}) to be able to include conformality loss via mechanism 3), while preserving all of the above mentioned tested features.

\section {A New Ansatz}\label{mall}

The simplest way to allow for the additional non-trivial UVFP in $\beta$-function of Eq. (\ref{unmod}), and thus anticipate for the `fixed point merger', is to include a term $\sim \gamma^2$ in the numerator of the $\beta$-function. The denominator is also modified due to the reasons we explain below. The modified all order $\beta$-function, in the notation introduced in the beginning of the previous section, is therefore
\be
\label{modif}  
	\beta(a) = -\frac{2}{3} a^2 \cdot \frac{\overline{\beta}_0(X) - X \,
	\gamma(a)+ r(X) \gamma^2(a)}{1- \frac{2aC_2(G)}{N_c}\left( 1+ \frac{6\overline{\beta}_0'(X)}{\overline{\beta}_0(X)}\right)+2Xa^2(1-2a)^{-1}},
\ee
where $\overline{\beta}_0(X)$ and $\overline{\beta}_0^\prime(X)$ are as defined earlier and we introduced unknown function $r(X)$ which will be specified below.

It is worth emphasizing that at no stage so far we have taken the large $N_c$ limit but we have simply rewritten everything using new variables and in Eq. (\ref{modif}) added a $\gamma^2$ term compared to Eq. (\ref{unmod}). By construction, everything at this stage is valid for any $N_c$.  In the terminology used in \cite{Kaplan:2009kr}, our $\beta$-function proposal, when applied to SU(3) with fundamental representation, corresponds to the QCD$^*=$QCD, i.e. the evolution of the gauge coupling has an { \it additional} unstable UVFP at the strong coupling, as we discussed in the introduction; in other words, within this approach the only relevant operators we consider at strong coupling are kinetic terms for the fermions and the gauge fields.

Before proceeding with the more detailed analysis of (\ref{modif}), let us briefly comment on its general form. The denominator has been chosen in order to reproduce the two-loop result upon expanding to ${\mathcal{O}}(a^3)$ and to match with super Yang--Mills in the large $N_c$-limit when the fermion content of the theory consists of a single Weyl fermion in the adjoint representation. However, our subsequent results are not dependent on the denominator, so we expect them to be more general. In fact, one could imagine to write the numerator in (\ref{modif}) as a more general expansion
\be
	\overline{\beta}_0(X)-q(X)\gamma(a)+r(X)\gamma^2(a)+s(X)a^2+\cdots,
	\label{general_numerator}
\ee 
with the omitted terms containing even higher terms in $\gamma$ and $a$. With the choice of the denominator as in (\ref{modif}), $q(X)=X$ is fixed by matching on the two-loop perturbative $\beta$-function. Since we aim to study situation where $\beta$-function has at most two zeros, we neglect any higher order contribution in $\gamma$ than the ones explicitly present in (\ref{general_numerator}).  Since the value of $\gamma(a^\ast)$ at the fixed point, as determined by setting the expression (\ref{general_numerator}) to zero, should be scheme-independent, it cannot depend explicitly on the value of the coupling and hence in our approach we have $s(X)\equiv 0$ and no higher order contributions proportional to higher powers of $a$ can appear either.   

As we mentioned above, the last term in the denominator allows us to reproduce the exact $\beta$-function of supersymmetric Yang--Mills theory in the limit of infinite number of colors assuming that the scheme is the same. Actually, it is straightforward to modify the denominator to provide this matching at any value of $N_c$, but since our interest is in nonsupersymmetric gauge theories and our results will be independent of the precise form of the denominator, we choose not to pursue this detail here. We are satisfied with the simple result that our $\beta$-function is compatible with the exact equivalence theorems in the literature \cite{Armoni:2003gp}.

Hence, with these remarks in mind we now return to the form (\ref{modif}) and aim to determine the function $r(X)$. We will first do it in the limit $N_c\rightarrow\infty$,  and then consider ${\mathcal{O}}(1/N_c)$ corrections in a later subsection.

\subsection{Large $N_c$ limit}
\label{large}

In the large $N_c$ limit we will apply the holographic expectation \cite{Kaplan:2009kr, Klebanov:1999tb} that operator dimensions of the quark mass operators at the two fixed points satisfy
\be
	\Delta_+ + \Delta_- = d = 4\,,
\ee 
which translates for the anomalous dimensions into equation
\be
\label{constraint}
	\gamma_1 + \gamma_2 = 2.
\ee 
Note, that this implies that fixed point merger, a point where conformality is lost, will occur at $\gamma_1=\gamma_2=1$.

Setting the numerator of the $\beta$-function (\ref{modif}) to zero and solving the resulting quadratic equation, with the constraint Eq. (\ref{constraint}) for the two roots, we find that
\be
	r(X)=X/2.
\label{rfunction}	
\ee
This fixes the final unknown coefficient of the $\beta$-function in Eq. (\ref{modif}).

Let us take a closer look at the terms in the numerator of $\beta$-functions in Eqs. (\ref{unmod}) and (\ref{modif}) containing the $\overline{\beta}_0(X)$ coefficient,
\be
	&&\overline{\beta}_0(X) -X \,\gamma \hspace{20mm} \text {All order $\beta$-function}\label{AOBF}\\
	&&\overline{\beta}_0(X) + \frac{X}{2} \,\gamma(\gamma-2) \hspace{9mm}  \text {Modified all order $\beta$-function } \label{MAOBF}.
\ee 
Comparing these, we observe the following: 

\begin{enumerate}
{ \item The modified $\beta$-function, in the large $N_c$ limit, unambiguously predicts the lower end of the conformal window. For a given theory, we have to solve for $X$ by setting the numerator of modified beta-function, Eq. (\ref{MAOBF}), equal to zero under the condition $\gamma=1$  on the basis of the constraint (\ref{constraint}) from holography. Therefore, by construction, the maximum value of the anomalous dimension at the IRFP is $\gamma^{\ast\rm{max}}=1$. In contrast, with (\ref{AOBF}) the lower end of the conformal window is not absolutely predicted but only bounded by demanding the absence of negative norm states in conformal field theory which corresponds to setting $\gamma=2$ in Eq.(\ref{AOBF}).  Of course, actual size of the conformal window may be smaller as chiral symmetry breaking could be triggered already for values $\gamma< 2$. Nevertheless, in (\ref{AOBF}) nothing forbids to have, at the fixed point, $\gamma^{\ast}>1$ assuming that given YM theory is conformal in the infrared. As an example, take an SU(3) gauge theory with two Dirac flavors in two-index symmetric (sextet) representation. Assuming this theory achieved an IRFP, $\beta$-function in Eq. (\ref{unmod}) predicts that $\gamma^{\ast}=1.3$ \cite{Sannino:2008ha}. We keep in mind, that our result $\gamma^{\ast\rm{max}}=1$ is valid only in the large $N_c$ limit. We will see that it will be possible to have $\gamma^{\ast\rm{max}}>1$ in a modified $\beta$-function (\ref{modif}) once we include ${\mathcal{O}}(1/N_c)$ corrections.}

{\item In both $\beta$-functions, upper end of the conformal window corresponds to $\gamma=0$ which translates to $\beta_0=0$ in Eqs. (\ref{AOBF}) and (\ref{MAOBF}). Notice also that the second value of the anomalous dimension satisfying $\beta_0=0$ in Eq. (\ref{MAOBF}) is $\gamma=2$ which coincides with the maximum value generally allowed in conformal field theory. }

{\item The perturbative IRFP analysis of \cite{Ryttov:2007cx} carried out for (unmodified) $\beta$-function remains valid for (\ref{modif}) since the anomalous dimension at the IRFP is small, $\gamma\ll 1$.}

{\item Pure YM ($N_f=0$) prediction from (\ref{modif}) is the same as from (\ref{unmod}) and therefore comparison with the lattice data performed for this case in \cite{Ryttov:2007cx} also holds for $\beta$-function in (\ref{modif}).}

{\item As already stated, similarly to the case of \cite{Ryttov:2007cx} we reproduce the exact NSVZ $\beta$-function of super Yang--Mills. Using (\ref{rfunction}) in (\ref{modif}) and considering one Weyl fermion in the adjoint representation, the resulting $\beta$-function is equated with the known SYM $\beta$. Solving for $\gamma$ leads to the NSVZ result and furthermore shows that with our $\beta$-function the quantity $a(Q)\langle\bar{\psi}\psi\rangle_Q$ is renormalization group invariant to all orders. This in turns establishes the equivalence between our scheme and the one used by NSVZ. Requiring this matching was the origin of the choice for the last term in the denominator of \ref{modif}. Recall, however, that the results concerning the size of the conformal window in this section (and in next sections) are blind to the precise form of the denominator in (\ref{modif}).}

\end{enumerate}

All that being said, let us now turn to the numerical predictions for the size of the conformal window with the new $\beta$-function ansatz. To determine the boundary of the conformal window corresponding to the transition from infrared conformal phase to chiral symmetry broken phase (i.e. the lower end of the conformal window) we set the expression in Eq. (\ref{MAOBF})  to zero and use that at the critical point  $\gamma_1=\gamma_2=1$. This leads to the prediction that, in the large $N_c$ limit, the lower end of the conformal window is at
\be
\label{nice}
	X_{\rm{min}}=\frac{22C_2(G)}{5N_c}=\frac{22}{5}.
\ee
where in the final step we used $G_2(G)=N_c$. For the SU$(N_c)$ gauge theory with the fundamental quarks this predicts that $x_{\rm{min}}=22/5$ in the large $N_c$ limit. Note that for fundamental fermions this implies also that we have $N_f\rightarrow\infty$, i.e. the Veneziano limit. We extrapolate further on this result in the left panel of Fig. \ref{upper} for the SU(3) theory with fundamental matter. There, above $x=11/2=5.5$  asymptotic freedom is lost (corresponding to the upper end of the conformal window),  while below $x=22/5=4.4$ the two zeros for $\gamma$ become complex. We therefore observe that for the SU(3), our extrapolation from large $N_c$ result gives the critical number of flavors $N_f^{cr}=13.2$ below which the conformality is lost. This disagrees with the latest lattice result that the lower end of the conformal window occurs for $8 \leq N_f \leq 12$ \cite{Appelquist:2007hu}.

Obvious reason for this disagreement is in the fact that Eq. (\ref{constraint}) was motivated in \cite{Kaplan:2009kr} by the holographic considerations where on the four dimensional field theory side one usually considers the large $N_c$ limit. Hence, by construction, we expect the results of this section to be valid in the large $N_c$ limit. Then for fundamental fermions in the Veneziano limit $x_{\rm{min}}=22/5$ is a prediction. Unfortunately we are not aware of any lattice simulations for SU($N_c$) gauge theory in the Veneziano limit to confront with our predictions.

It is therefore clear that we should expect important ${\mathcal{O}}(1/N_c)$ corrections to (\ref{constraint}) which, in turn, would modify the $r(X)$ coefficient of the $\gamma^2$ term. We will include these effects in the following subsection by means of a phenomenological modifying parameter $\epsilon$ as $\gamma_1 + \gamma_2 = 2 +\epsilon$.

\begin{figure}[htb]
\centerline{\includegraphics[width=.9\linewidth]{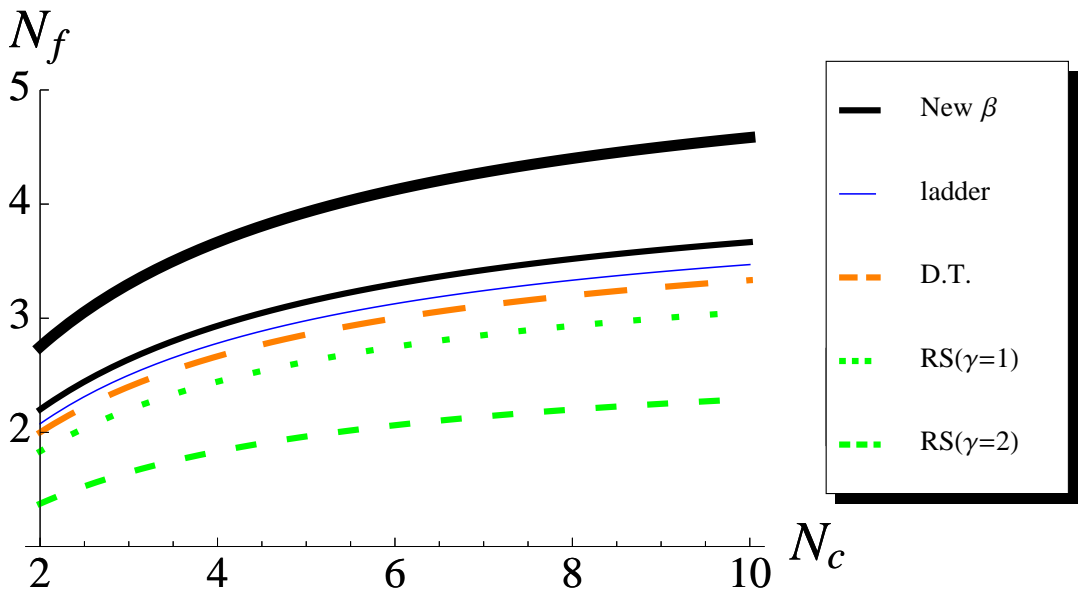}}
\caption{Size of the conformal window determined using different methods discussed in the text with $N_f$ fermions in 2-index symmetric representation of SU($N_c$). The topmost thick solid curve denotes the upper boundary of the conformal window, i.e. loss of asymptotic freedom. For the lower boundary the lower solid curve is the prediction from our $\beta$-function proposal, Eq.(\ref{nice}), the long dashed curve is the prediction from \cite{Poppitz:2009uq} and the curve with shorter dashes correspond to the prediction from \cite{Ryttov:2007cx} for $\gamma=2$ (the dotted curve shows the corresponding result for $\gamma=1$). The thin line between the solid and long-dashed curves corresponds to the ladder approximation.} 
\label{2s_cw}
\end{figure}

Before that, let us also perform some additional numerical tests for our proposal. Consider first calculating the lower end of the conformal window in terms of critical number of Dirac fermions $N_f^{cr}$ from (\ref{nice}) for fermions in the two-index symmetric (2S) representation of the SU($N_c$) gauge group. As discussed above, we keep in mind that for low number of colors we expect corrections to (\ref{constraint}) so that only the result in the large $N_c$ limit should be trustworthy and others should be interpreted with care. Results are presented in Fig.\ref{2s_cw} together with results corresponding to the $\beta$-function ansatz of \cite{Ryttov:2007cx} and with recent results \cite{Poppitz:2009uq} obtained using deformation theory to establish the role of topological excitations in generating a mass gap in a gauge theory. Specifically these latter results apply to the gauge theory quantized on a {\bf $R^3 \times S^1$ } where an appropriate matter content is introduced to preserve the (approximate) center symmetry at any radius. The estimate for the lower end of the conformal window is then obtained using the expectation that for conformal theories the topological excitations become irrelevant as the size of {\bf $S^1$} is increased, while for confining theories instead, they become more relevant. In the figure, also the result from the ladder approximation is shown. 

We will return to the comparisons between different methods in Sec. \ref{corrections} after estimating the ${\mathcal{O}}(1/N_c)$ corrections. At this stage we simply observe from Fig.\ref{2s_cw} that the predictions from Ryttov-Sannino $\beta$-function as well as those from the deformation theory method are consistently lower than ours for the minimum $N_f$ for which the conformal window is reached. 
As to the confinement mechanism from \cite{Poppitz:2009uq}, for the SU($N$)-2S it is due to magnetic bions; quantum mechanically stable topological composites carrying a net magnetic charge.  

The SU(2)-2S gauge theory with two Dirac flavors has been investigated on the lattice in \cite{Hietanen:2008mr,Catterall:2007yx}. These studies indicate that either the theory is very near the IRFP or the IRFP is already reached.  Additionally, lattice results in \cite{Shamir:2008pb,DeGrand:2008kx} suggest that the SU(3)-2S theory with two Dirac flavors may already achieved an IRFP. We also note that the phenomenology of a walking technicolor theory with two flavors either in two-index symmetric representation of SU(2) \cite{Antipin:2009ks} or SU(3) \cite{Belyaev:2008yj} has been studied recently. 

\subsection {Away from the large $N_c$ limit \label{corrections}}

Now, let us include the ${\mathcal{O}}(1/N_c)$ effects by means of introducing parameter $\epsilon$ as  $\gamma_1 + \gamma_2 = 2 +\epsilon$. Repeating then the exercise of Sec. \ref{large}, we arrive at the following coefficient of the $\gamma^2$ term 
\be
	r(X)=\frac{X}{2+\epsilon},
\ee
and, thus, zero of the numerator of the new beta-function will occur at:
\be\label{modnice}
	\overline{\beta}_0(X) + X \,
	\gamma(\frac{\gamma}{2+\epsilon}-1)=0.
\ee 
or, when writing $\overline{\beta}_0(X)$ explicitly,
\be
	11\frac{C_2(G)}{N_c}+X\left[\gamma\left(\frac{\gamma}{2+\epsilon}-1\right)-2\right]=0,
\ee
where again for SU($N_c$) gauge group $G_2(G)/N_c=1$.
\begin{figure}[htb]
\centering
\includegraphics[width=3.0in]{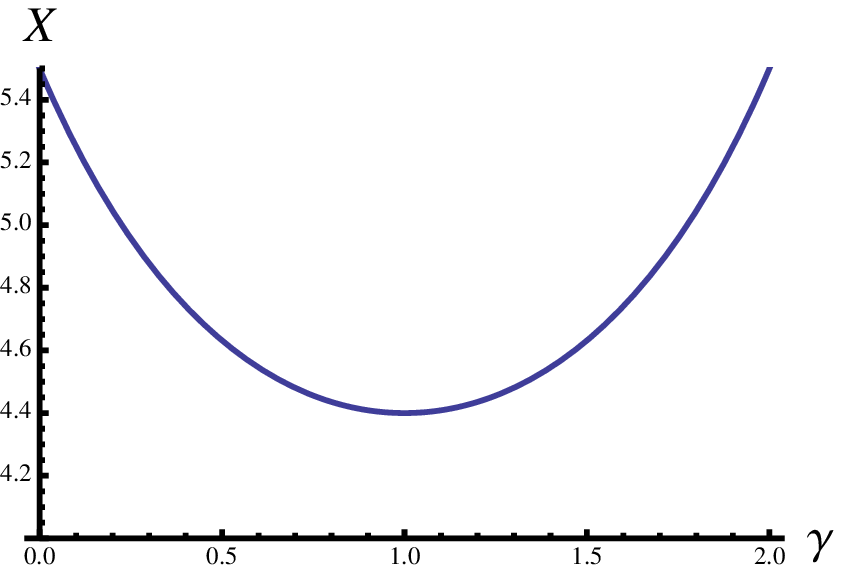} \includegraphics[width=3.0in]{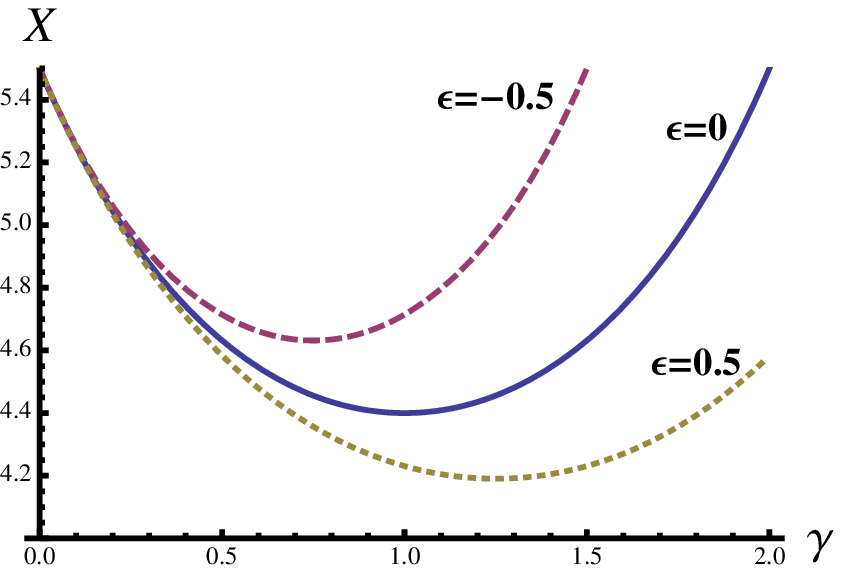}
\caption{Left: The line $X(\gamma)$ along which $\beta(a)=0$. Above $X=11/2=5.5$  asymptotic freedom is lost while below $X=22/5=4.4$ no real solutions exist. Right: The effects of $1/N_c$ corrections. Solid curve is the same as in the left panel. Dashed curve corresponds to $\epsilon=-0.5$ and dotted to $\epsilon= 0.5$}
\label{upper}
\end{figure}

In the right panel of Fig. \ref{upper} we plot the solution $X(\gamma)$ of this equation for the SU(3) fundamental with the illustrative values $\epsilon=\pm 0.5$ and  $\epsilon=0$. The latter coincides with the curve in the left panel of Fig. \ref{upper}. We observe that fixed point merger happens at $\gamma= 0.75$ and $\gamma= 1.25$ for $\epsilon=-0.5$ and $\epsilon=0.5$, respectively. The upper end of the conformal window at X=11/2=5.5 corresponding to $\gamma_1= 0$ remains unchanged. As a qualitative effect we notice that the size of the conformal window increases (decreases) with positive (negative) $\epsilon$ and for positive $\epsilon$ the anomalous dimension at the UVFP will exceed the unitarity limit $\gamma\le 2$ before the IRFP merges with the free UVFP, i.e. before the other solution reaches $\gamma=0$ value. 

In other words, as we  decrease $\gamma$ at the IRFP (by increasing $\epsilon$ with fixed $X$ or vice versa) there is a point where UVFP disappears due to violation of the unitarity bound. Thus, past this point our picture becomes qualitatively similar to the one predicted by Ryttov-Sannino $\beta$-function. The value of  $\gamma$ at the IRFP below which the UVFP ceases to exist is given by $\gamma=\epsilon$. 

\begin{figure}[htb]
\centering
\includegraphics[width=2.0in]{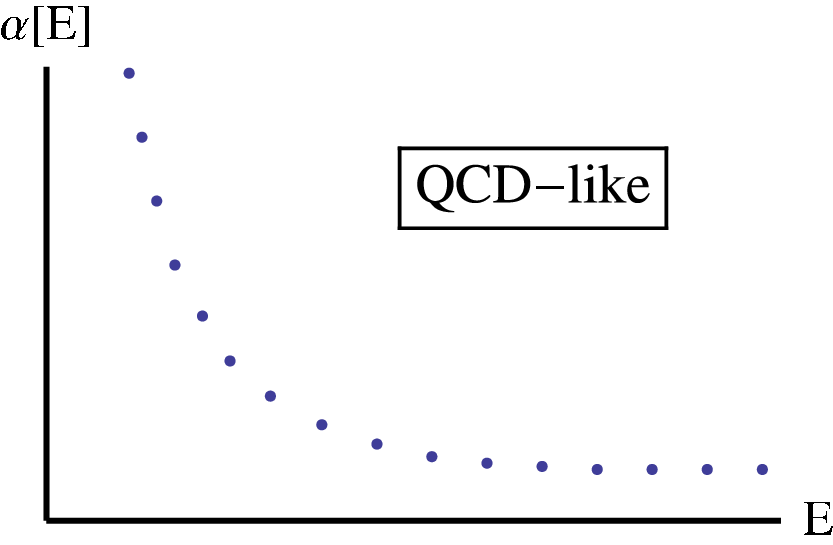} \includegraphics[width=2.0in]{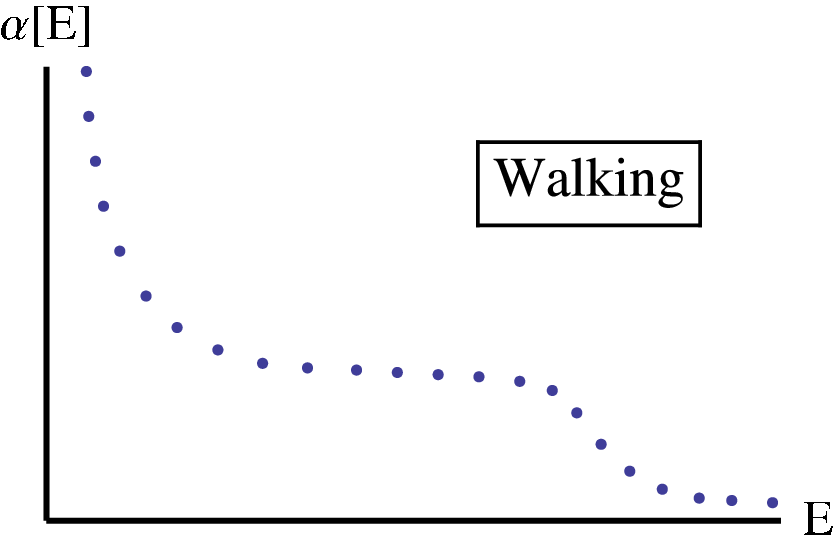} \includegraphics[width=2.0in]{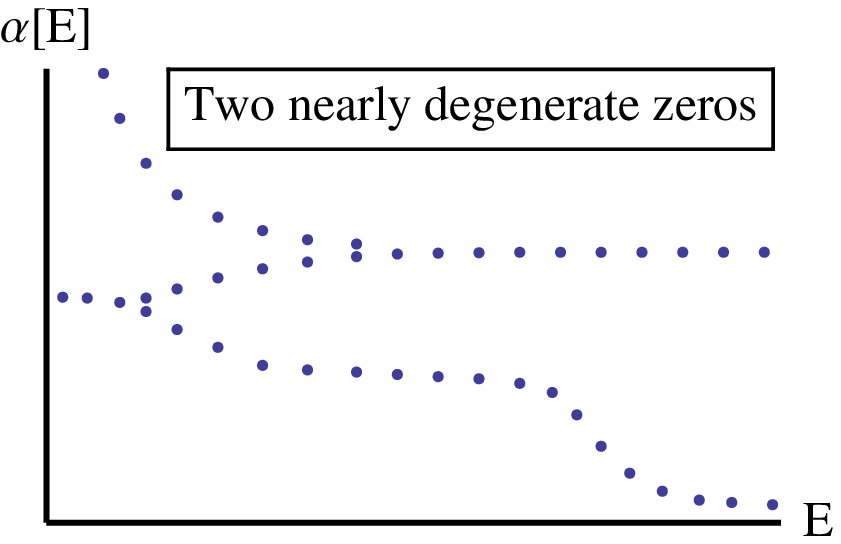}\\
\includegraphics[width=2.0in]{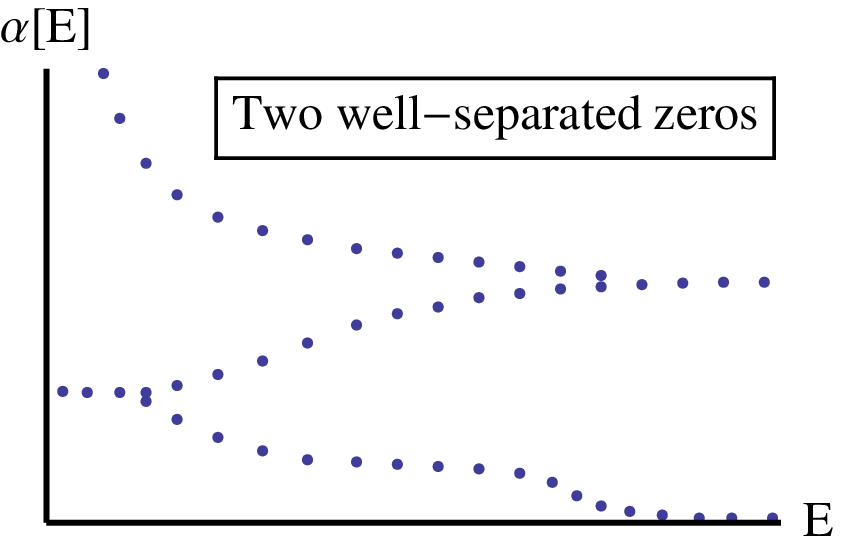} \includegraphics[width=2.0in]{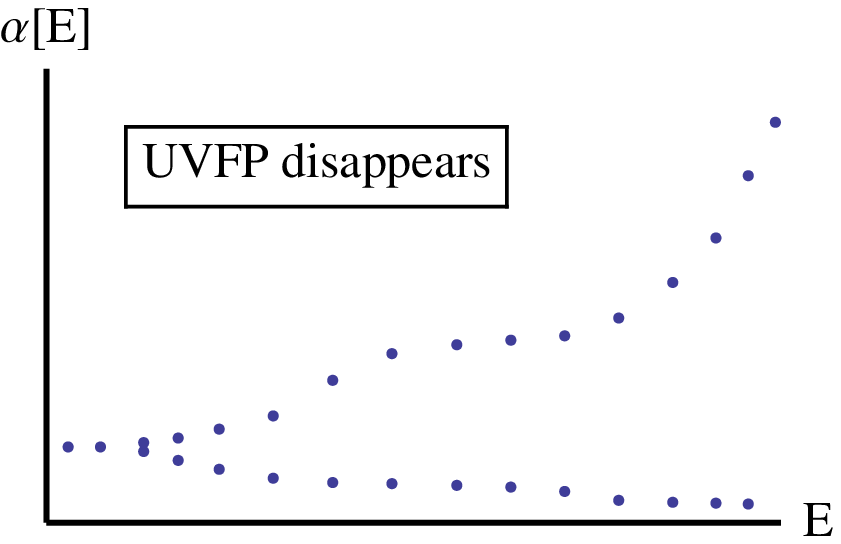} \includegraphics[width=2.0in]{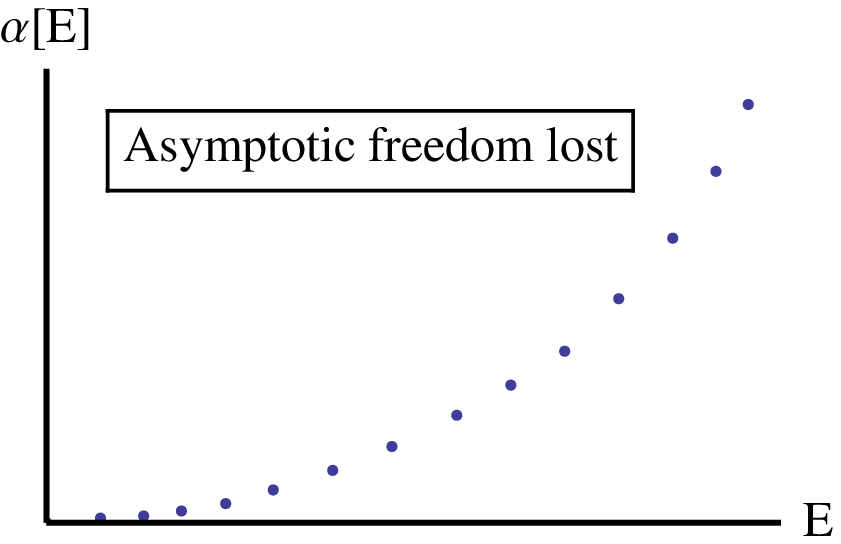}
\caption{Running of the gauge coupling as a function of energy scale implied by our $\beta$-function proposal.}
\label{final}
\end{figure}

We illustrate schematically these behaviors by plotting the running of the gauge coupling as a function of energy scale in Fig. \ref{final}. We go from left to right in Fig. \ref{final} as we increase $X$. Initially, we are well below the lower end of the conformal window and have QCD-like behavior of the coupling constant. When we approach the non-trivial zero of the $\beta$-function, coupling runs slowly and we have a walking-like behavior. Increasing X further we cross past the  
merger and two nearly degenerate zeros of the $\beta$-function appear and, with increasing $X$, they separate further away until at the UVFP the unitarity bound is violated. At this point we have only the non-trivial IRFP. Finally this IRFP merges with the free UVFP as we exit conformal window at the upper end. 

Above discussion modifies slightly if $\epsilon<0$. The difference with the above case is that both IRFP and UVFP remain present until the point where asymptotic freedom is lost. At this point we reach the maximal value for anomalous dimension at the unstable UVFP, $\gamma_{UVFP}^{\rm{max}}=2-\epsilon$, while IRFP merges with the free UVFP. In terms of the graphs in Fig. \ref{final}, the phase labeled "UVFP disappears" will be absent.

Above, we have treated $\epsilon$ as a phenomenological parameter whose value is expected to be ${\mathcal{O}}(1)$. To obtain semi-quantitative estimates for conformal windows away from large $N_c$ limit we will set $\epsilon=1$. 

As a heuristic discussion how this value might emerge, we detour a little via the AdS/CFT correspondence starting from the well known result that for a given mass $m$ of a scalar in $AdS_{d+1}$ there are two solutions, $\Delta_{\pm}$, for the dimension of the corresponding operators in the conformal field theory on the boundary
\be
	\Delta_{\pm}= \frac{d}{2}\pm\sqrt{\frac{d^2}{4}+m^2}\label{KW}.
\ee
We now recall from\cite{Klebanov:1999tb} (see also \cite{Breitenlohner:1982jf}) that for:
\be\label{KW1}
	-\frac{d^2}{4} < m^2 < -\frac{d^2}{4}+1.
\ee
there are two AdS-invariant quantizations corresponding to the $z^{\Delta_+}$ and $z^{\Delta_-}$ asymptotic behavior of the scalar wave functions near the AdS boundary. Either solution can be realized but the action is finite only for $\Delta > d/2$ (corresponding to $\gamma<d/2-1$). But this bound was shown to be relaxed by adding appropriate boundary terms to the action leading to the weaker condition $\Delta> (d-2)/2$ (corresponding to $\gamma<d/2$) which coincides with the unitarity bound on the dimension of a scalar operator in $d$ dimensions. For the mass range in Eq. (\ref{KW1}), this unitarity bound allows both $\Delta_{\pm}$ solutions while for larger $m^2$ only $\Delta_+$ is permitted. 

Since our $\beta$-function is not limited only to the conformal theories, we cannot take the results of \cite{Klebanov:1999tb} too literally. Within the language of our proposal, we imagine that in the region where our $\beta$-function has two non-trivial zeros we expect the dimensions of the quark mass operators at the fixed points to be bounded by the unitarity bound only (i.e. $\gamma<2)$, while in the region of only one zero the stronger bound ($\gamma < 1$) applies. We notice that the picture we just described is featured for the $\epsilon > 0$ case only; see the right panel of Fig.\ref{upper}, where two fixed points coexist until the UVFP violates the unitary bound and therefore disappears. 

This formulation also leads to an upper bound for $\epsilon$, since as the UVFP disappears due to the corresponding anomalous dimension becoming larger than two, the anomalous dimension at remaining IRFP has to be bounded by $\gamma\le 1$ as discussed above. This, in turn, leads to the upper bound $\epsilon < 1$. 

Combining these holographic results we now consider only $0\le \epsilon\le 1$ and use this bound on $\epsilon$ to correct the lower boundary of the conformal window determined in the previous subsection. Solving Eq. (\ref{modnice})  at the point where the UV and IR fixed points merge (i.e. where $\gamma_1=\gamma_2=1+\epsilon/2$) we obtain
\be\label{bound}
	X_{\rm{min}}=\frac{22C_2(G)}{N_c(5+\epsilon/2)}=\frac{22}{5+\epsilon/2},
\ee 
where the second equality again applies for SU($N_c$) gauge theory. Consequently, we propose to use the upper bound on $\epsilon$ for an estimate of the lower end of the conformal window as follows: For SU($N_c$) with $\overline{\beta}_0(X)=11-2X\ge 0$, use $\epsilon_{\rm{max}}=1$ 
as an error estimate for finite $N_c$ corrections. This diagnostics leads to the $X_{\rm{min}}=4$ value predicted by Eq. (\ref{bound}).

Let us apply this result first for SU(3) with fundamental fermions. Using $x_{\rm{min}}=X_{\rm{min}}=4$ this translates to  $N_{f,{\rm{min}}}=12$, in agreement with the recent lattice results.

Then let us consider various values of $N_c$ and also higher representations in addition to the fundamental one. We collect the results into the Table \ref{corr}.
\begin{table}
\caption{Lower end of the conformal window using procedure outlined in text for fundamental (F), 2-index (anti)symmetric (2AS) 2S and adjoint (A) representations.}
\label{corr}
\begin{center}
\begin{tabular}{|c|c|c|c|c|c|}
\hline
$N_c$ & $N_{f,{\rm{min}}}$ Fund. & $N_{f,{\rm{min}}}$ 2AS & $N_{f,{\rm{min}}}$ 2S & $N_{f,{\rm{min}}}$ A.\\
\hline
2          & 8           & -       &  2	     & 2 \\	
3          & 12         & 12    &  2.4     & 2 \\
4          & 16         & 8      &  2.67   & 2 \\
5          & 20         & 6.67 &  2.86   & 2 \\
6          & 24         & 6      &  3        & 2 \\
10        & 40         & 5      &  3.33   & 2 \\
$\to\infty$& $4N_c$ &4      & 4       & 2 \\
\hline

\end{tabular}
\end{center}
\end{table}
Interestingly, our results are consistent with present lattice results: The SU(3) gauge theory with 12 fundamental flavors appears to possess a fixed point \cite{Appelquist:2007hu}, as seems to be the case also for SU(2) with two adjoint flavors \cite{Hietanen:2008mr}. In contrast, the most recent study of SU(3) with two flavors in the two-index symmetric representation seems to be just outside the conformal window \cite{DeGrand:2008kx} which is also suggested by our bound in Table \ref{corr}. Remarkably, our results for two-indexed and adjoint representations match exactly on the corresponding predictions from deformation theory \cite{Poppitz:2009uq}. Even more intriguingly, taking into account the recent refinements within this framework \cite{Poppitz:2009tw}, also the numbers in the column corresponding to the fundamental representation coincide exactly 

Of course any relation between our $\beta$-function ansatz and some underlying microscopic dynamics is pure speculation; nevertheless the coincidence with \cite{Poppitz:2009uq} although unexpected is temptingly systematic. Our results also agree with the ones in \cite{Armoni:2009jn}.

As a further comparison, we can estimate the values of the gauge coupling at the lower end of the conformal window and compare with the corresponding critical values of the gauge coupling predicted by ladder approximation \cite{Appelquist:1988yc,Cohen:1988sq}. Here one needs to resort to perturbative formulas for the anomalous dimension. Using the one-loop formula (\ref{gamma_pert}) we find 
\be\label{coupling}
	\alpha_c&=&\frac{\pi}{3C_2(R)} \hspace{10mm} \text {\,\,Ladder approximation },\\
	\alpha_c&=&\frac{2 \gamma^\ast}{3C_2(R)} \hspace{10mm} \text {\,\,Both all order beta-functions },
\ee
where $\gamma^\ast$ stands for the anomalous dimension evaluated at the IRFP.
From these equations we notice that at the critical point the value of the coupling predicted by the modified $\beta$-function is smaller than the ladder result; recall that the maximum value of the anomalous dimension at the IRFP, for finite values of $\epsilon$, was 1.5. Needless to say, usage of one-loop result for $\gamma\sim 1$ is hardly justified and should not be taken too seriously. One can try to improve by using two-loop formula for $\gamma$, and this generally has the effect of making the numerical value of the coupling $\alpha_c$ smaller; the same effect also happens with the Ryttov-Sannino $\beta$-function ansatz \cite{FS_private}. This is a nice feature also in light of recent lattice simulations which, in the cases of fundamental \cite{Appelquist:2007hu} and higher representations \cite{Hietanen:2008mr,Shamir:2008pb}, systematically seem to observe a fixed point at relatively small value of the coupling. All this seems to imply a consistent picture that even if the numerical value of the coupling is small (in some scheme), there can be large nonperturbative effects. 

\section{Conclusions and further prospects}
\label{conclusion}

In this paper we exploited recently proposed mechanism of IR and UV fixed point annihilation as a key difference between supersymmetric and non-supersymmetric YM gauge theories. We incorporated this mechanism of fixed point merging into a proposal for a nonperturbative $\beta$-function whose functional form is similar to the Ryttov-Sannino $\beta$-function. Our conjectured form of the $\beta$-function was shown to exhibit rich dynamics. 

We summarized the dynamical picture implied by our new $\beta$-function as an energy behavior of the gauge coupling  and analyzed the conformal window predicted by our proposal and confronted it with the existing lattice data. Further improvements in lattice results are needed in order to support/reject our $\beta$-function more conclusively. We also compared our results with results from Ryttov-Sannino $\beta$-function \cite{Ryttov:2007cx} as well as with results of \cite{Poppitz:2009uq}. As a speculative note, the agreements and disagreements between these three approaches suggest that there might be important differences in how conformality is lost in supersymmetric versus non-supersymmetric gauge theories

The estimates for the conformal windows in gauge theories based on Sp($2N_c$) and SO($N_c$) groups have appeared in\cite{Sannino:2009aw} using the Ryttov-Sannino $\beta$-function and in \cite{Golkar:2009aq} using the deformation theory methods. Since our proposal can be equivalently well applied there, we checked these as well. Recall that our result for the lower boundary of the conformal window was 
\be
	X_{\rm{min}}=\frac{22C_2(G)}{N_c(5+\epsilon/2)},
\ee
which implies that (taking $\epsilon=1$)
\be
	N_{f,{\rm{min}}}=4\frac{C_2(G)}{2T(R)}.
\ee

For Sp($2N_c$) the required group theory factors are $C_2(G)=N_c+1$ and $T(F)=1/2$, $T(A)=N_c+1$, $T(2AS)=N_c-1$ while for SO($N_c$) these are, $C_2(G)=N_c-2$ and $T(F)=1$, $T(A)=N_c-2$, $T(2S)=N_c+2$. The notation is the same which we have used in earlier sections, namely F, A, 2S and 2AS denote fundamental, adjoint, two-index symmetric and antisymmetric representations. Note that the adjoint representation coincides for Sp($2N_c$) with two-index symmetric and for SO($N_c$) with two-index antisymmetric representation for all $N_c$. The results are summarized in Table \ref{spso}. Comparing with corresponding results compactly tabulated in \cite{Golkar:2009aq} we conclude that our results match closely again with the deformation theory for adjoint and two-index representations while disagree more with the fundamental. 

\begin{table}[h]
\caption{Summary of the results for the lower boundary of the conformal window for Sp($2N_c$) (left) and SO($N_c$) (right) gauge theories for fundamental (F), adjoint (A) and two-index (2S or 2AS) representations.}
\label{spso}\flushleft
\begin{minipage}[t]{.5\linewidth}
	\begin{tabular}{|c|c|c|c|}
	\hline
	$N_c$ & $N_{f,{\rm{min}}}$, F & $N_{f,{\rm{min}}}$, A & $N_{f,{\rm{min}}}$, 2AS \\
	\hline
	2	   & 12                              & 2                               & 6 \\
	3         & 16                              & 2                               & 4 \\
	4         & 20                              & 2                               & 3.4\\
	5         & 24                              & 2                               & 3 \\
	10       & 44                              & 2                               & 2.4 \\
	$\to\infty$ & $4N_c$          &2 				   & 2 \\
	\hline
	\end{tabular}
\end{minipage}\begin{minipage}[t]{.5\linewidth}
	\begin{tabular}{|c|c|c|c|}
	\hline
	$N_c$ & $N_{f,{\rm{min}}}$, F & $N_{f,{\rm{min}}}$, A & $N_{f,{\rm{min}}}$, 2S \\
	\hline
	6	   & 8                                & 2                               & 1 \\
	7         & 10                              & 2                               & 1.1 \\
	8         & 12                              & 2                               & 1.2 \\
	9         & 14                              & 2                               & 1.3 \\
	10       & 16                              & 2                               & 1.33 \\
	$\to \infty$  & $2N_c$             & 2                               & 2 \\
	\hline
	\end{tabular}
\end{minipage}
\end{table}

Finally, some further developments of the original proposal in \cite{Ryttov:2007cx} have appeared in the literature: inclusion of multiple fermion representations was considered in \cite{Ryttov:2009yw} and finite quark mass and flavor dependence in \cite{Dietrich:2009ns}. Similar extensions could be carried out also for the proposal we have presented in this paper, but we will leave these for future work. As a yet another direction for future investigation using these phenomenological $\beta$-function ans\"atze would be the construction of the 5D holographic dual gravity theory coupled to the dilaton and the axion using the method outlined in \cite{Gursoy:2007cb}. In this approach, dilaton potential is shown to be in one-to-one correspondence with the exact beta-function of gauge theory, and its knowledge determines the full structure of the vacuum solution; however, including the anomalous dimension of the quark mass operator explicitly present in our approach would require some further efforts. 

\acknowledgments
We thank A.\,Armoni, H.\,Gies,  F.\,Sannino and M.\,Unsal for useful and insightful correspondence.

\end{document}